\documentclass[aps,prl,epsfig,showpacs,amsmath,twocolumn]{revtex4}

\usepackage{graphicx}
\newcommand{\ba}{\begin{eqnarray}}
\newcommand{\ea}{\end{eqnarray}}
\newcommand{\ban}{\begin{eqnarray*}}
\newcommand{\ean}{\end{eqnarray*}}
\newcommand{\bsub}{\begin{subequations}}
\newcommand{\esub}{\end{subequations}}

\newcommand{\nc}{\newcommand}
\nc{\Id}{{\mathchoice {\rm 1\mskip-4mu l} {\rm 1\mskip-4mu l}
{\rm 1\mskip-4.5mu l} {\rm 1\mskip-5mu l}}}

\begin{document}

\title{
Symmetries and Supersymmetries of the Dirac Hamiltonian with\\ 
Axially-Deformed Scalar and Vector Potentials}
\author{A. Leviatan}

\affiliation{
Racah Institute of Physics, The Hebrew University, 
Jerusalem 91904, Israel}
\date{\today}

\begin{abstract}
We consider several classes of symmetries 
of the Dirac Hamiltonian in 3+1 dimensions, with axially-deformed scalar 
and vector potentials. 
The symmetries include the known pseudospin and spin 
limits and additional symmetries which occur when 
the potentials depend on different variables. 
Supersymmetries are observed within each class and 
the corresponding charges are identified.
\end{abstract}

\maketitle

The Dirac equation plays a key role in microscopic descriptions of 
many-fermion systems, employing covariant density functional theory and the 
relativistic mean field approach. 
The relevant mean-field potentials are of Coulomb vector type in atoms, 
and a mixture of Lorentz vector and scalar potentials in nuclei and 
hadrons~\cite{wal86}. 
Recently, symmetries of Dirac Hamiltonians with 
such mixed Lorentz structure have been shown to be relevant for explaining 
the observed degeneracies of certain shell-model orbitals in nuclei 
(``pseudospin doublets'')~\cite{gino97}, 
and the absence of quark spin-orbit splitting 
(``spin doublets'') \cite{page01}, as observed in heavy-light quark 
mesons. Supersymmetric patterns have been 
identified in specific limits of such 
spherical potentials~\cite{suku85,lev04}. 
In the present Letter we further explore classes of symmetries 
and supersymmetries when these potentials are axially-deformed. 
Such a study is significant in view of the fact that mean-field 
Hamiltonians often break the rotational symmetry. Cylindrical 
geometries are relevant to a number of problems, including electron 
channeling in crystals, structure of 
axially-deformed nuclei and quark confinement in spheroidal flux-tubes. 

The Dirac Hamiltonian, $H$, for a fermion of mass~$M$ 
moving in external scalar, $V_S$, and vector,
$V_V$, potentials is given by 
$H = \mbox{\boldmath $\hat{\alpha}\cdot \hat{p}$}
+ \hat{\beta} (M  + V_S) + V_V$~\cite{Thaller92}. 
When the potentials are axially-symmetric, {\it i.e.}, independent 
of the azimuthal angle $\phi$, $V_{S,V}=V_{S,V}(\rho,z)\,$, 
$\rho = \sqrt{x^2+y^2}$, then 
the $z$-component of the angular momentum 
operator, $\hat{J}_z$, commutes with 
$H$ and its half-integer eigenvalues $\Omega$ are used to label the 
Dirac wave functions 
$\Psi = 
\left ( g^{+} e^{- i\phi/2},\,
g^{-} e^{i\phi/2},\,
if^{+} e^{- i\phi/2},\,
if^{-} e^{i\phi/2}
\right )e^{i\Omega\phi}$. 
Here $g^{\pm}\equiv g^{\pm}(\rho,z)$ and $
f^{\pm}\equiv f^{\pm}(\rho,z)$ are the radial wave functions of the 
upper and lower components, respectively. 
Henceforth, such a wave function will be denoted by 
$\Psi_{\Omega}:\, \{g^{+},g^{-},f^{+},f^{-}\}$. 
The potentials enter the Dirac equation through the combinations 
\bsub
\ba
\label{A}
A(\rho,z) &=& E + M + V_S(\rho,z) - V_V(\rho,z) \\
\label{B}
B(\rho,z) &=& E - M - V_S(\rho,z) - V_V(\rho,z) ~.
\ea
\label{AB}
\esub 
For each solution with $\Omega>0$, 
there is a degenerate time-reversed solution 
with $-\Omega <0$, hence, 
we confine the discussion to solutions with $\Omega>0$. 
Of particular interest 
are bound Dirac states with $\vert E\vert < M$ and normalizable 
wave functions 
in potentials satisfying 
$\rho V_S(\rho,z),\rho V_V(\rho,z) \rightarrow 0$ 
for $\rho\rightarrow 0$ and 
$V_S(\rho,z),V_V(\rho,z) \rightarrow 0$ 
for $\rho\rightarrow \infty$ or $z\rightarrow\pm\infty$. 
The boundary conditions imply that the radial wave functions 
fall off exponentially for large distances and behave as 
a power law for $\rho\rightarrow 0$. 
Furthermore, for $z=0$ and $\rho\rightarrow \infty$, 
$f^{-}/g^{+} \propto (M-E)>0$ and $g^{-}/f^{+} \propto (M+E) > 0$, 
while for $z=0$ and $\rho\rightarrow 0$, 
$f^{-}/g^{+}\propto B(0)\rho$ and 
$g^{-}/f^{+}\propto -A(0)\rho$.
These properties have important implications for the 
structure of radial nodes. In particular, it follows that 
for potentials with the indicated asymptotic behaviour and 
$A(0),\,B(0)>0$, as encountered in nuclei, 
a necessary condition for a nodeless bound 
eigenstate of a Dirac Hamiltonian is 
\ba
g^{-} =0 \;\; {\rm or}\;\; f^{+} =0 ~.
\label{nodeless}
\ea
The Dirac equation, $H\Psi = E\Psi$, leads to a set of four coupled 
partial differential equations involving the radial wave functions. 
Their solutions are greatly simplified in the presence of symmetries. 
We now discuss four classes of relativistic symmetries 
and possible supersymmetries within each class.

The symmetry of class I, referred to as pseudospin symmetry, 
occurs when the sum of the scalar and vector 
potentials is a constant, 
$V_{S}(\rho,z) + V_{V}(\rho,z) = \Delta_0$. 
The symmetry generators, ${\hat{\tilde {S}}}_{i}$, 
commute with the Dirac Hamiltonian 
 and span an SU(2) algebra~\cite{bell75,ginolev98}
\ba
{\hat{\tilde {S}}}_{i} = 
\left (
\begin{array}{cc}
 U_p\, \hat{s}_i\, U_p &  0 \\
0 & \hat{s}_{i}
\end{array}
\right ) 
\quad i =x,y,z 
\quad U_p = \, \frac{\mbox{\boldmath $\sigma\cdot p$}}{p} ~.
\label{pSgen}
\ea
Here 
${\hat s}_{i} = \sigma_{i}/2$ are the usual spin operators, 
defined in terms of Pauli matrices. 
The Dirac eigenfunctions in the pseudospin limit satisfy
\ba
{\hat{\tilde {S}}}_{z}
\Psi^{(\tilde{\mu})}_{\Omega} &=&
\tilde{\mu}\Psi^{(\tilde{\mu})}_{\Omega}\qquad
\;\tilde{\mu} = \pm 1/2
\ea
and form degenerate $SU(2)$ doublets. 
Their wave functions have 
been shown to be of the form~\cite{gino05}
\bsub
\ba
\label{Psi1ps}
\Psi^{(-1/2)}_{\Omega_1=\tilde{\Lambda}-1/2}: &\; 
\left \{\, g^{+},\,-g,\,0,\,f\,\right \} ~,\\
\label{Psi2ps}
\Psi^{(1/2)}_{\Omega_2=\tilde{\Lambda}+1/2}: &\; 
\left \{\, g,\, g^{-},\,f,\, 0\,\right \} ~,
\ea
\label{wfps}
\esub
where $\tilde{\Lambda} = \Omega-\tilde{\mu}\geq 0$ 
is the eigenvalue of $\hat{J}_z - {\hat{\tilde {S}}}_{z}$.
The relativistic pseudospin symmetry has 
been tested in numerous realistic mean field calculations 
of nuclei and were found to be obeyed to a good approximation, 
especially for doublets near the Fermi surface~\cite{ginlev04,gino05}. 
The dominant upper components of the states in Eq.~(\ref{wfps}),  
involving $g^{+}$ and $g^{-}$, correspond to non-relativistic 
pseudospin doublets with asymptotic (Nilsson) quantum numbers 
$[N,n_3,\Lambda]\Omega=\Lambda+1/2$ and 
$[N,n_3,\Lambda +2]\Omega=\Lambda+3/2$, respectively. 
The doublet is expressed in terms of the 
pseudo-orbital angular momentum projection, $\tilde{\Lambda}=\Lambda+1$, 
which is added to the pseudospin projection, $\tilde{\mu}=\pm 1/2$, 
to form doublet states with  $\Omega=\tilde{\Lambda}\pm 1/2$. 
Such doublets play a crucial role in explaining features of 
deformed nuclei, including superdeformation and identical 
bands~\cite{gino05,bohr82}.

The symmetry of class II, referred to as spin symmetry, occurs 
when the difference of the scalar and vector potentials is a constant,  
$V_{S}(\rho,z) - V_{V}(\rho,z) = \Xi_0$. 
The symmetry group is again $SU(2)$ and its generators~\cite{bell75}
\ba
{\hat{S}}_{i} = 
\left (
\begin{array}{cc}
\hat{s}_{i} & 0 \\
0 & U_p\, \hat{s}_i\, U_p
\end{array}
\right ) 
\quad i =x,y,z ~
\label{Spgen}
\ea
commute with the Dirac Hamiltonian. 
The Dirac eigenfunctions in the spin limit satisfy
\ba
\hat{S}_{z}
\Psi^{(\mu)}_{\Omega} &=&
\mu\,\Psi^{(\mu)}_{\Omega}\qquad
\;\mu = \pm 1/2
\ea
and form degenerate $SU(2)$ doublets. Their wave functions 
are of the form~\cite{gino05}
\bsub
\ba
\label{Psi1sp}
\Psi^{(1/2)}_{\Omega_1=\Lambda+1/2}: &\;& 
\left \{\, g,\,0,\, f,\, f^{-}\,\right \} ~,\\
\label{Psi2sp}
\Psi^{(-1/2)}_{\Omega_2=\Lambda-1/2}: &\;& 
\left \{\, 0,\,g,\, f^{+},\,-f\,\right \} ~,
\ea
\label{wfsp}
\esub
where $\Lambda = \Omega -\mu\geq 0$ is the eigenvalue of 
$\hat{J}_z - \hat{S}_{z}$.
The upper components of the two states in Eq.~(\ref{wfsp}) 
form the usual non-relativistic spin doublet with a common radial wave 
function, $g$, an orbital angular momentum projection, 
$\Lambda$, and two spin orientations 
$\Omega = \Lambda\pm 1/2$. 
The relativistic spin symmetry has been shown to be relevant to the 
structure of heavy-light quark mesons~\cite{page01}. 

The Dirac Hamiltonian has additional symmetries when the scalar and 
vector potentials depend on different variables. 
The symmetry of class III occurs when  
the potentials are of the form
$V_{S}=V_{S}(z)$ and $V_{V}=V_{V}(\rho)$. 
In this case, the Dirac Hamiltonian commutes with the following 
Hermitian operator 
\ba
\hat{R}_{z} &=& \left [\,M+V_{S}(z)\,\right ]\hat{\beta}\,\hat{\Sigma}_3 
+\gamma_{5}\hat{p}_z ~,
\label{Rz}
\ea 
where $\hat{\Sigma}_i = \left ({\sigma_i\atop 0}{0\atop \sigma_i}\right )$. 
The Dirac eigenfunctions satisfy
\ba
\hat{R}_z\,\Psi^{(\epsilon)}_{\Omega} &=& 
\epsilon\,\Psi^{(\epsilon)}_{\Omega} ~.
\label{RzPsi}
\ea
A separation of variables is possible by choosing the Dirac 
wave function in the form 
\ba
\Psi^{(\epsilon)}_{\Omega}: &\,&  
\left \{\, 
u_1h_{+},\,
u_2h_{-},\,
u_1h_{-},\,
-u_2h_{+}\,\right \}/\sqrt{\rho} ~,\quad
\label{PsiRz}
\ea
where $u_{i}\equiv u_{i}(\rho)$, $h_{\pm}\equiv h_{\pm}(z)$ 
and, for simplicity, we have omitted the label $\epsilon$ from these 
wave functions. 
The Dirac equation then reduces to a set of two coupled first-order 
ordinary differential equations in the variable $\rho$, 
\bsub
\ba
\left [  d/d\rho - \Omega/\rho  \right ] u_{1}(\rho)
-\left [ E - V_{V}(\rho) +\epsilon \right ] u_{2}(\rho) &=& 0 
\;\;\;\;\;\qquad \\ 
\left [  d/d\rho + \Omega/ \rho  \right ] u_{2}(\rho)
+\left [ E - V_{V}(\rho) -\epsilon \right ] u_{1}(\rho) &=& 0 
\ea
\label{rhoeq}
\esub
and a separate set in the variable $z$ 
\bsub
\ba
\left [M + V_{S}(z) + d/d z \right ]h_{2}(z) 
&=& \epsilon\, h_{1}(z)\\
\left [ M + V_{S}(z) - d/d z \right ]h_{1}(z) 
&=& \epsilon\, h_{2}(z)
\ea
\label{zeq}
\esub
where $h_{\pm}(z)=h_{2}(z)\pm h_{1}(z)$. 
The separation constant,~$\epsilon$, 
plays the role of a mass for the transverse motion 
and is determined from imposed boundary conditions. 
A~special case within 
the symmetry class III, with $V_S(z)=0$ and  
$\epsilon = \pm\sqrt{M^2 + p_{z}^2}$, was considered 
for electron channeling in 
crystals~\cite{channeling}. 
For $V_{S}(z)=0$, $\hat{R}_z$ of Eq.~(\ref{Rz}), reduces to the transverse 
polarization operator relevant to studies of synchrotron radiation 
in storage rings and QED processes in magnetic flux tubes 
({\it e.g.}, $e^{+}e^{-}$ production 
and Bremsstrahlung)~\cite{skar96}. 
\begin{table*}[t]
\centering
\label{Tab1}
\caption{{\it Conserved, anticommuting operators for Dirac Hamiltonians 
$(H)$ exhibiting a supersymmetric structure.}} 
\begin{tabular}{lccc}
\hline\hline
\noalign{\smallskip}
\hspace{1cm} SUSY & $\hat{R}$ 
& $\hat{B}$ & $\hat{B}^2=f(H)$ \\
\noalign{\smallskip}\hline\noalign{\smallskip}
$V_{S}(\rho,z) + V_{V}(\rho,z) = \Delta_0\quad$
& $\hat{\tilde {S}}_{z}$ (\ref{pSgen}) & 
 $2(M+\Delta_0 - H)\hat{\tilde {S}}_{x}$ & 
 $( M+\Delta_0 - H)^2$ \\ 
$V_{S}(\rho,z) - V_{V}(\rho,z) = \Xi_0\quad$
& $\hat{S}_{z}$ (\ref{Spgen})          & 
 $ 2(\,M+\Xi_0 + H)\,\hat{S}_{x}$ &
 $ (M+\Xi_0 + H)^2 $ \\ 
$V_S = V_{S}(z)$, $V_V = \frac{\alpha_{V}}{\rho}$ 
& $\hat{R}_{z}$ (\ref{Rz}) &
$\quad\hat{\beta}\,\hat{\Sigma}_3
\{i\hat{J}_{z}\gamma_5 [\,H -  \hat{\Sigma}_3\hat{R}_z ]
-\frac{\alpha_{V}}{\rho}
(\mbox{\boldmath $\hat{\Sigma}\cdot\rho$})\,\hat{R}_z \} $ &
$ \quad\hat{J}_{z}^2 ( H^2 - \hat{R}_{z}^2) 
+ \alpha_{V}^2 \hat{R}_{z}^2\quad $ \\ 
$V_S = \frac{\alpha_{S}}{\rho}$, $V_V = V_{V}(z)$ 
& $\hat{R}_{\rho}$ (\ref{Rrho}) &            
$\quad\hat{\Sigma}_3\{i\hat{J}_{z}\gamma_5 [M - \hat{\Sigma}_3 
\hat{R}_{\rho}] -\frac{\alpha_{S}}{\rho} 
(\mbox{\boldmath $\hat{\Sigma}\cdot\rho$}) 
\hat{\beta}\hat{R}_{\rho}\}$ &
$\quad\hat{J}_{z}^2 
(\hat{R}_{\rho}^2 - M^2) 
+ \alpha_{S}^2 \hat{R}_{\rho}^2$ \\
\noalign{\smallskip}\hline\hline
\end{tabular}\\
\end{table*}

The symmetry of class IV occurs when the potentials are of the form 
$V_{S}=V_{S}(\rho)$ and $V_{V}=V_{V}(z)$. 
In this case, the following Hermitian operator 
\ba
\hat{R}_{\rho} &=&
[\,M+V_{S}(\rho)\,]\hat{\Sigma}_3 
-i\hat{\beta}\,\gamma_{5}
(\,\mbox{\boldmath ${\hat{\Sigma}}\times\hat{p}$}\,)_{3}
\label{Rrho}
\ea
commutes with the Dirac Hamiltonian and the 
Dirac eigenfunctions satisfy
\ba
\hat{R}_{\rho}\,\Psi^{(\tilde{\epsilon})}_{\Omega} &=& 
\tilde{\epsilon}\,\Psi^{(\tilde{\epsilon})}_{\Omega} ~.
\label{RrhoPsi}
\ea
Again, a separation of variables is possible with the choice of 
wave function,
\ba
\Psi^{(\tilde{\epsilon})}_{\Omega}: &\,&  
\left \{\, 
\xi_1w_{+},\,
-i\xi_2w_{-},\,
i\xi_1w_{-},\,
-\xi_2w_{+}\,\right \}/\sqrt{\rho} ~,\quad
\label{PsiRrho}
\ea
where $\xi_{i}\equiv \xi_{i}(\rho)$ and $w_{\pm}\equiv w_{\pm}(z)$. 
The Dirac equation then reduces to a set 
of ordinary differential equations in the variable $\rho$, 
\bsub
\ba
\left [ d /d\rho - \Omega/\rho\right ] \xi_{1}(\rho)
-\left [ \tilde{\epsilon} + M + V_{S}(\rho)\, \right ] 
\xi_{2}(\rho) &=& 0 
\;\;\;\;\;\qquad \\ 
\left [ d /d\rho + \Omega/\rho\right ] \xi_{2}(\rho)
+\left [ \tilde{\epsilon} - M  - V_{S}(\rho)\, \right ] \xi_{1}(\rho) &=& 0
\ea
\label{rhoeq2}
\esub
and a separate set in the variable $z$ 
\bsub
\ba
\left [E - V_{V}(z) - id/d z \right ]w_{2}(z) 
&=& \tilde{\epsilon}\, w_{1}(z)\\
\left [ E - V_{V}(z) + id/d z \right ]w_{1}(z) 
&=& \tilde{\epsilon}\, w_{2}(z)
\ea
\label{zeq2}
\esub
where $w_{\pm}(z)=w_{2}(z)\pm w_{1}(z)$. 
The quantum number,~$\tilde{\epsilon}$, plays the role of an energy for 
the transverse motion. 
A particular selection of potentials within the symmetry class~IV 
was encountered in the study of the Schwinger mechanism for 
particle-production in a strong confined 
field  ($V_{V}(z) =\alpha_{V}z$) \cite{wang88,gat92}, 
$q\bar{q}$ pair-creation in a flux tube 
$(V_{S}(\rho)=0,\, \tilde{\epsilon} = 
\pm\sqrt{M^2 + k^2}\,)$~\cite{pavel91}, 
and the canonical quantization in cylindrical geometry of 
a free Dirac field $(V_{S}(\rho)=V_{V}(z)=0,\, E^2 = 
M^2 + k^2 + p_{z}^2)$ \cite{cooper06}. 

Dirac Hamiltonians with selected external fields 
are known to be supersymmetric~\cite{lev04,suku85,Thaller92,junker96}. 
It is, therefore, natural to inquire whether 
a supersymmetric structure can develop 
within each of the above symmetry classes.
The essential ingredients of supersymmetric quantum 
mechanics~\cite{junker96} 
are the supersymmetric Hamiltonian, ${\cal H}$,  
and charges $Q_{+}$, $Q_{-} = Q_{+}^{\dagger}$, 
which generate the supersymmetry (SUSY) algebra
$[{\cal H},Q_{\pm}] = \{Q_{\pm},Q_{\pm}\}=0$, 
$\{Q_{-},Q_{+}\}= {\cal H}$. 
Accompanying this set is an Hermitian $Z_2$-grading 
operator satisfying $[{\cal H},{\cal P}] = \{Q_{\pm}, {\cal P}\}=0$ 
and ${\cal P}^2 = \Id$. 
The $+1$ and $-1$ eigenspaces of ${\cal P}$ define the ``positive-parity'', 
$H_{+}$, and ``negative-parity'', $H_{-}$, sectors of the spectrum, with 
eigenvectors $\Psi^{(+)}$ and $\Psi^{(-)}$, respectively. 
The SUSY algebra imply that if $\Psi^{(+)}$ is 
an eigenstate of ${\cal H}$, then also 
$\Psi^{(-)}=Q_{-}\Psi^{(+)}$ is an 
eigenstate of ${\cal H}$ with the same energy, 
unless $Q_{-}\Psi^{(+)}$ vanishes or produces an unphysical state, 
({\it e.g.}, non-normalizable). 
The resulting spectrum consists of pairwise degenerate levels 
with a non-degenerate single state (the ground state) 
in one sector when the supersymmetry 
is exact. If all states are pairwise degenerate, 
the supersymmetry is said to be broken. 
Typical spectra for good and broken SUSY are shown in Fig.~1.
\begin{figure}[b]
  \includegraphics[height=.2\textheight,angle=-90]{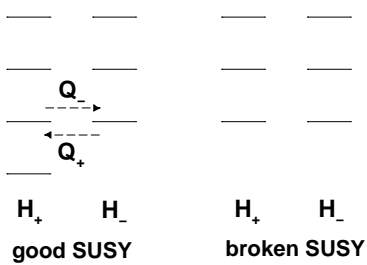}
  \caption{Typical spectra of good and broken SUSY. The operators $Q_{-}$ 
and $Q_{+}$ connect degenerate states in the $H_{+}$ and $H_{-}$ 
sectors.}
\end{figure}
\begin{figure*}
\begin{center}
\leavevmode
\includegraphics[width=0.80\linewidth]{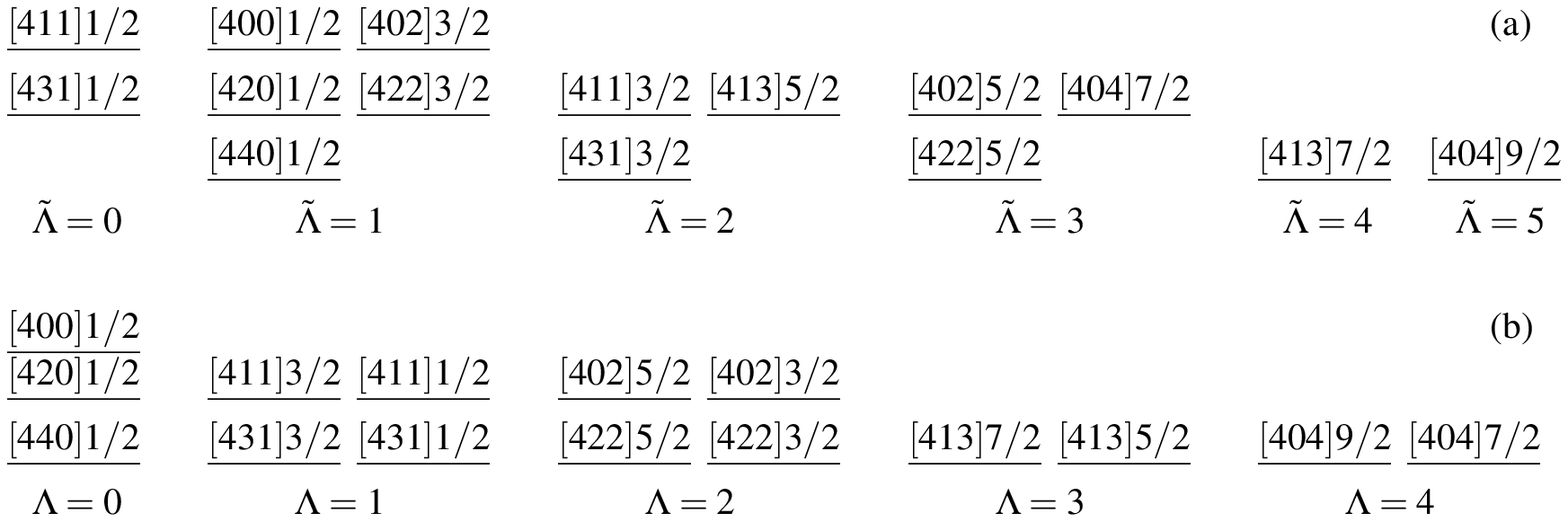}\\
\caption{
Grouping of deformed shell-model states
$[N=4,n_3,\Lambda]\Omega$, exhibiting 
a pattern of 
(a) good SUSY, relevant 
to the pseudospin symmetry limit, and (b) broken SUSY, relevant 
to the spin symmetry limit. $N$ and $n_3$ are harmonic oscillator 
quantum numbers. $\Lambda$ ($\tilde{\Lambda}$)  
is the orbital (pseudo-orbital) angular momentum projection along the symmetry 
$z$-axis.}
\end{center}
\end{figure*}
Degenerate doublets, signaling a supersymmetric structure, 
can emerge in a quantum system with a Hamiltonian $H$, from the existence 
of two Hermitian, conserved and anticommuting operators, $\hat{R}$ and 
$\hat{B}$
\ba
[ H, \hat{R}] = [H, \hat{B}] = \{\hat{R},\hat{B}\} =0~.  
\label{HRB}
\ea
The operator $\hat{R}$ has non-zero eigenvalues, $r$, 
which come in pairs of opposite signs. 
$\hat{B}^2 = \hat{B}^{\dagger}\hat{B} = f(H)$, is a function 
of the Hamiltonian. A $Z_2$-grading operator, 
${\cal P}_r = \hat{R}/\vert r \vert$, and Hermitian supercharges 
$Q_1 = \hat{B}$, 
$Q_2 = iQ_1{\cal P}_{r}$
can now be constructed. 
The triad of operators 
$Q_{\pm} = (Q_1 \pm iQ_2)/2$ and ${\cal H}=Q_{1}^2=f(H)$ 
form the standard SUSY algebra. 
In the present analysis, $f(H)$ is a quadratic function of the Dirac 
Hamiltonian, $H$, and the relevant $\hat{R}$ and $\hat{B}$ operators 
are listed in Table~I. 

In the pseudospin symmetry limit, the relevant operator $\hat{B}$, 
connects the doublet states of Eq.~(\ref{wfps}). 
The spectrum, for each $\tilde{\Lambda}\neq 0$, 
consists of twin towers of pairwise degenerate pseudospin doublet states,  
with $\Omega_1=\tilde{\Lambda}-1/2$ and $\Omega_2=\tilde{\Lambda}+1/2$, 
and an additional non-degenerate 
nodeless state at the bottom of the $\Omega_1=\tilde{\Lambda}-1/2$ tower. 
Such nodeless states correspond in the non-relativistic nuclear deformed 
shell-model to the ``intruder'' states, 
$[N,n_3,\Lambda=N-n_3]\Omega=\Lambda+1/2$, 
which, empirically, are found not to be part of a doublet~\cite{bohr82}. 
The latter property follows from the fact that 
a nodeless bound Dirac state satisfies the criteria 
of Eq.~(\ref{nodeless}), hence has a wave function as in 
Eq.~(\ref{Psi1ps}) with $g^{+},\, g,\, f\neq 0$ and 
$f/g^{+} >0$. Its pseudospin partner state has a wave function as 
in Eq.~(\ref{Psi2ps}). The radial components satisfy 
$Bg^{-} = [B - 2(\tilde{\Lambda}/\rho)f/g^{+}]g^{+}$, 
where $B$ is defined in Eq.~(\ref{B}). 
This relation is satisfied, to a good approximation, 
for mean-field potentials relevant to nuclei, and the r.h.s. 
is non-zero and, consequently, $g^{-}\neq 0$. If so, then 
the partner state (\ref{Psi2ps}) 
is also nodeless, but it cannot be a bound eigenstate 
since its radial components do not fulfill 
the condition of Eq.~(\ref{nodeless}). 
Altogether, the ensemble of Dirac states with $\Omega_2-\Omega_1=1$ exhibits 
a supersymmetric pattern of good SUSY, as illustrated 
in Fig.~(2a).

In the spin symmetry limit, the relevant operator $\hat{B}$ connects 
the doublet states of Eq.~(\ref{wfsp}). 
The spectrum, for each $\Lambda\neq 0$, 
consists of twin towers of pairwise degenerate spin-doublet states 
with $\Omega_1=\Lambda-1/2$ and $\Omega_2=\Lambda+1/2$. 
None of these towers have a 
single non-degenerate state. 
This follows from the fact that, in view of Eq.~(\ref{nodeless}),  
a nodeless bound state has a wave function as in Eq.~(\ref{Psi1sp}) 
with $g,\, f,\, f^{-}\neq 0$ and $g/f^{-} >0$. 
Its spin partner has a wave function as in Eq.~(\ref{Psi2sp}). 
The radial components satisfy 
$Af^{+} = [ A - 2(\Lambda/\rho)g/f^{-} ] f^{-}$, where $A$ is defined in 
Eq.~(\ref{A}). For relevant potentials the r.h.s. of this relation 
can vanish, hence $f^{+}$ has a node. 
Therefore, the spin-partner of a nodeless state 
is not nodeless and can be a bound eigenstate, since the restrictions of 
Eq.~(\ref{nodeless}) do not apply. 
Altogether, the ensemble of Dirac states with 
$\Omega_2-\Omega_1=-1$ exhibits a supersymmetric pattern 
of broken SUSY, as illustrated in Fig.~(2b). 

Within the symmetry class~III, a 
supersymmetry is obtained for 
$V_{V}(\rho) = \alpha_{V}/\rho$ and $V_{S}(z)$ arbitrary. 
The energy eigenvalues are 
$E^{(\epsilon)}_{n_{\rho},\Omega} = 
|\epsilon|/\sqrt{
1 + \alpha_{V}^2/(n_{\rho}+\gamma)^2}\;$ 
$(n_{\rho} =0,1,2,\ldots)$, 
with $\gamma = \sqrt{\Omega^2 - \alpha_{V}^2}$.
From Eqs.~(\ref{zeq}) we see that if $[h_{1}(z), h_{2}(z)]$ are 
solutions with $\epsilon>0$, then $[h_{1}(z), -h_{2}(z)]$ are solutions 
with $-\epsilon<0$. Accordingly, the doublet wave functions 
are as in Eq.~(\ref{PsiRz}), 
with the replacements, $u_{i}\mapsto u_{i}^{(\epsilon)}(\rho)$ 
for $\Psi^{(\epsilon)}_{n_{\rho},\Omega}$, and
$u_{i}\mapsto u_{i}^{(-\epsilon)}(\rho)$, $h_{\pm}\mapsto -h_{\mp}(z)$ 
for $\Psi^{(-\epsilon)}_{n_{\rho},\Omega}$. 
For $n_{\rho}\geq 1$, the states 
$\Psi^{(\pm\epsilon)}_{n_{\rho},\Omega}$ are degenerate. 
For $n_{\rho}=0$ only one state is an acceptable solution, 
which has $\epsilon>0$ (assuming $\alpha_V < 0$) 
and is annihilated by the relevant operator $\hat{B}$. 
For each $\Omega$ and $\epsilon$ the spectrum 
resembles a supersymmetric pattern of good SUSY, with the towers $H_{+}$ 
($H_{-}$) of Fig.~1 corresponding to states with 
$\epsilon>0$ ($\epsilon<0$).

Within the symmetry class~IV, a 
supersymmetry is obtained for 
$V_{S}(\rho) = \alpha_{S}/\rho$ $(\alpha_{S} <0)$ 
and $V_{V}(z)$ arbitrary.   
The allowed values are
$\tilde{\epsilon} = 
\pm M\sqrt{
1 -\alpha_{S}^2/(n_{\rho}+\tilde{\gamma})^2}$
$(n_{\rho}=0,1,2,\ldots)$, 
where $\tilde{\gamma} = \sqrt{\Omega^2 + \alpha_{S}^2}$. 
From Eqs.~(\ref{zeq2}) we see that 
if $[w_{1}(z),w_{2}(z)]$ are solutions with 
$\tilde{\epsilon}>0$, then $[w_{1}(z),-w_{2}(z)]$ are solutions 
with $-\tilde{\epsilon}<0$ and the same energy, $E$. 
Accordingly, the doublet wave-functions 
are as in Eq.~(\ref{PsiRrho}), 
with the replacements, $\xi_{i}\mapsto \xi_{i}^{(\tilde{\epsilon})}(\rho)$ 
for $\Psi^{(\tilde{\epsilon})}_{n_{\rho},\Omega}$, and
$\xi_{i}\mapsto \xi_{i}^{(-\tilde{\epsilon})}(\rho)$, 
$w_{\pm}\mapsto -w_{\mp}(z)$ 
for $\Psi^{(-\tilde{\epsilon})}_{n_{\rho},\Omega}$. 
For $n_{\rho}\geq 1$ the states 
$\Psi^{(\pm\tilde{\epsilon})}_{n_{\rho},\Omega}$ are degenerate. 
For $n_{\rho}=0$ only one state, with $\tilde{\epsilon}>0$,  
is an acceptable solution, which is annihilated by the relevant 
operator $\hat{B}$. 
Again,  for each $\Omega$ and $\tilde{\epsilon}$ the resulting spectrum 
resembles a supersymmetric pattern of good SUSY.

In summary, we have considered classes of symmetries and 
related supersymmetries of Dirac Hamiltonians with 
cylindrically-deformed scalar and vector potentials.
The symmetries arise when the potentials obey a constraint on their 
sum or difference, or when they depend on different variables. 
The known pseudospin and spin symmetry limits are by themselves 
supersymmetric. Additional supersymmetries arise when one of 
the potentials has a $1/\rho$ dependence and the second 
potential depends on $z$. 
It is gratifying to note that some of the indicated (super)symmetries 
are manifested empirically, to a good approximation, in 
physical dynamical systems. 

This work was initiated during a Sabbatical stay at LANL and 
is supported by the Israel Science Foundation.
Discussions with J.N. Ginocchio are acknowledged.

\end{document}